\documentclass[a4paper,12pt]{article}
\usepackage{amsfonts,graphicx,slashed}
\newcommand{\ket}[1]{\mbox{$|#1\rangle$}}
\newcommand{\bra}[1]{\mbox{$\langle #1|$}}
\newcommand{\braket}[2]{\mbox{$\langle #1|#2\rangle$}}
\newcommand{\tr}{\mbox{tr }}

\begin{document}
\hfuzz=100pt
\title{{\LARGE \bf{Does black-hole evaporation imply that physics is non-unitary, and if so, what must the laws of physics look like? An Essay.}}}
\author{Steffen Gielen\footnote{Electronic address: sg452@damtp.cam.ac.uk}
\\
\\ D.A.M.T.P.,
\\ Cambridge University,
\\ Wilberforce Road,
\\ Cambridge CB3 0WA,
\\ U.K.
}

\maketitle

\begin{abstract}
Stephen Hawking's discovery of black hole evaporation had the remarkable consequence that information is destroyed by a black hole, which can only be accommodated by modifying the laws of quantum mechanics. Different attempts to evade the information loss paradox were subsequently suggested, apparently without a satisfactory resolution of the paradox. On the other hand, the attempting to include non-unitarity into quantum mechanics might lead to laws predicting observable consequences such as nonlocality or violation of energy-momentum conservation; but it may be possibly to circumvent these obstacles. Recent developments seem to require a different view on quantum gravity and suggest that ideas about locality in physics and Hawking's semi-classical approximation are misleading. An accurate description may show unitary evolution and no information loss after all.\end{abstract}
\eject

\tableofcontents

\section{Introduction}
Stephen Hawking's discovery that black holes radiate thermally has extremely striking consequences: It allows evolution of pure quantum states into mixed quantum states. The conclusion that has to be drawn from this ``information loss paradox" is that our present laws of quantum mechanics are insufficient for describing a process such as the formation and subsequent evaporation of a black hole, and that the possibility of non-unitary time evolution has to be introduced for an adequate description. Hawking's discovery and its consequences are explained in more detail in section two.
\\
\\However, since the framework of quantum mechanics seems to rest on unitarity, most physicists will tend to look for possible ways to get around such a drastic modification. These attempts are based on the fact that Hawking's calculation was done in an approximation where quantum fluctuations of the spacetime geometry are neglected, and so one might expect quantum effects to become important in an accurate description. It will be shown in section three that it seems impossible to find a consistent description in which pure states will always remain pure.
\\
\\In section four I shall analyse the question what non-unitary laws of physics could possibly look like in more detail. Inevitable violations of energy or momentum conservation, locality and causality, or Lorentz covariance were found in some models that allow non-unitarity in quantum mechanics. It also seems that Noether's theorem is inapplicable when time evolution of quantum states is non-unitary. Other authors came to different conclusions and were apparently able to construct models without these pathologies. I will discuss whether any satisfactory resolution of the debate has been reached.
\\
\\Stephen Hawking, who started the whole discussion, had his own viewpoint on how one should think about dynamic processes in curved spacetimes, due to the difficulties with defining local observables and time evolution in quantum gravity. He rejected a local description of quantum gravity, and the global description of quantum gravity led him to conclude that information loss was not a problem and that evolution in quantum mechanics was fully unitary.
\\There are other general arguments why nonlocal effects should become apparent in processes involving quantum gravity, possibly restoring unitarity, and why the arguments apparently ruling out unitary evolution may not be applicable to black-hole evaporation. These more recent ideas will be discussed in section five.
\\
\\In section six, the results will be summarised, with an outlook to future developments. The appendix contains a few calculations in detail that have been left out of the main part of this essay, and also a short introduction to the path integral approach to quantum gravity.
\\
\\I will use Planck units, so that $G=c=\hbar=1$. Also, there is no Boltzmann's constant.

\subsection{Carter-Penrose Diagrams}
Throughout this essay, Carter-Penrose diagrams will be used to illustrate the causal relations in a given spacetime. These diagrams are conventional in relativity, but perhaps less well known in general. To construct such a diagram, as first done by Carter in \cite{carter}, one chooses a (1+1)-dimensional submanifold of the spacetime. Since we will always basically consider an evaporating Schwarzschild black hole with spherical symmetry, we will choose a submanifold $\theta=\varphi={\rm const}$. One then goes to different coordinates which are of finite range, e.g. by the transformation $t=\tan T$ for time. Of course, now the metric will become infinite as these new coordinates approach their upper and lower limits. By applying a conformal transformation to metric, one can obtain a metric which is regular even as these limits are approached, so that it is possible to attach a ``boundary" to the spacetime. In the previous example, this boundary would consist of $T=\pm\frac{\pi}{2}$.
\\
\\While a conformal transformation changes scales and distances, it leaves the causal relations in the spacetime invariant, since the light cones of a metric $g$ and the rescaled version $\Omega^2 g$ are identical. Since any two-dimensional metric is conformal to a flat metric, by an appropriate choice of coordinates one can achieve that the rescaled metric is flat. Then a diagram of the (1+1)-dimensional new spacetime with its boundaries illustrates the causal relations of the original spacetime; at each point of this diagram one can draw light cones at 45 degrees which give the directions of light rays. The boundary ``points" (two-spheres in the four-dimensional spacetime) $\imath^+$ and $\imath^-$ are called ``future timelike infinity" and ``past timelike infinity"; timelike geodesics of the original spacetime can be extended to begin or end in these points in the new compact spacetime. Similarly, spacelike geodesics end in ``spacelike infinity" $\imath^0$. Light rays always begin and end in ``past null infinity" $\mathcal{J}^-$ and ``future null infinity" $\mathcal{J}^+$, respectively.
\\
\\It is conventional to draw wiggly lines for singularities and solid lines for event horizons and boundaries of spacetime. I use a dotted line for $r=0$ which corresponds to the origin in spherical polar coordinates, and is not to be confused with a boundary of spacetime. In Fig. 1, a light ray reaching $r=0$ would just be ``reflected" out to positive values of $r$ again. Spacelike infinity $\imath^0$ corresponds to $r=\infty$. The area covered by collapsing matter in Fig. 1 does not have any physical significance -- since scales change under conformal transformations, one could transform it into a different shape or area. However, the event horizon is always at 45 degrees. A particle that has passed it can not escape on a timelike curve which always has to be at less than 45 degrees to the vertical. In Fig. 4, the ``baby universe" can be reached by going through the singularity of the black hole.

\section{From Black Hole Evaporation to Non-Unitary Laws of Physics}

\subsection{Hawking Radiation and Information Loss}
By doing an analysis of quantum field theory at the event horizon, Stephen Hawking discovered in 1973 that a black hole emits radiation exactly like a blackbody at the Hawking temperature
\[T=\frac{\kappa}{2\pi},\]
where $\kappa$ is the surface gravity of the black hole \cite{hawking75}. For a Schwarzschild black hole, the total flux of radiation emitted per time is proportional to $M^2$, so that the time for a black hole to completely evaporate is proportional to $M^3$, where $M$ is the black hole mass.
\\
\\As a heuristic, but demonstrative, picture of this process, one might think of pair creation just outside the event horizon caused by vacuum fluctuations, so that one particle can escape to infinity while the other one falls into the hole, this one having negative energy so that total energy is conserved \cite{hawking75}. Although this particle appears to have negative energy relative to infinity, it can exist as a real particle with timelike momentum vector because the Killing vector which represents time translations is spacelike inside the horizon.
\\
\\Hawking derived this famous result in a ``semi-classical" approximation; he used quantum field theory on a curved background with a spacetime satisfying the classical Einstein equations. On a length scale as small as Planck size, quantum fluctuations of the metric itself are generally expected to cause the calculation to break down.
\\The notable fact about Hawking radiation is that it is exactly thermal and thus completely independent of the detailed structure of anything that collapsed to form the black hole. This is plausible since according to the uniqueness theorems (``no-hair theorems") of general relativity, the geometry outside a black hole is characterised only by the parameters mass, charge, and angular momentum.
\begin{figure}[h]
\begin{center}
\includegraphics[scale=0.5]{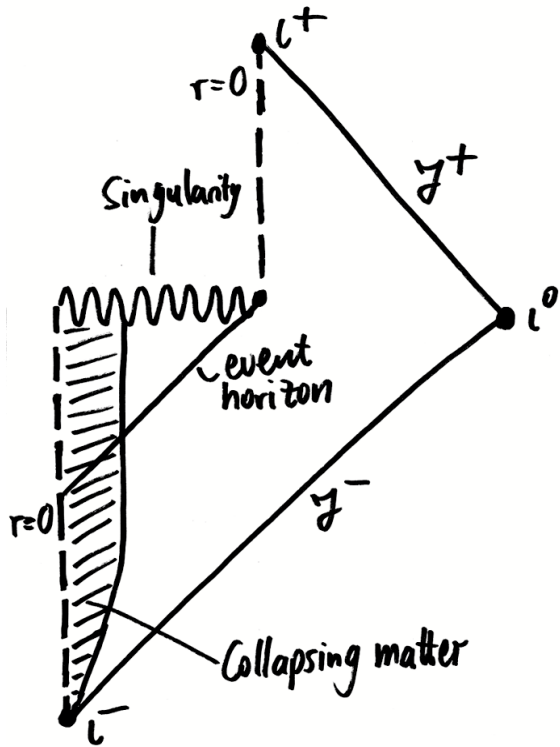}
\end{center}
\caption{Carter-Penrose diagram for gravitational collapse followed by evaporation of the black hole \cite{hawking75}}
\end{figure}
\\
\\This is the origin of the information loss paradox \cite{preskill}: A black hole will continue evaporating until it has disappeared, leaving only thermal radiation. In particular, a system that is in a pure quantum state might collapse to form a black hole; after the process of black hole formation and evaporation is completed, this system has evolved into a mixed state, which means that non-unitary evolution has occurred, in conflict with the laws of quantum mechanics. This is the paradox that will be discussed in this essay.
\\
\\Firstly, what does ``information loss" mean in this context? Given a quantum system in a pure state, there exist {\it non-degenerate} observables so that a measurement on the system will give a unique predictable result \cite{page}. (Of course, this is not true for general observables.) This is a property of pure states; for a mixed state there are no such observables. In this sense, complete information exists only for pure states.
Looking at the von Neumann entropy
\[S=-\tr(\rho\cdot\log\rho),\]
$S=0$ if and only if $\rho$ represents a pure state, since then the only eigenvalues of $\rho$ are one and zero. Information loss (or loss of quantum coherence) is then defined as an increase in the entropy $S$, which is exactly what happens in black hole evaporation. For any unitary operator $U$
\[S(U\rho U^{\dagger})=S(\rho),\]
since $S$ only depends on the eigenvalues of $\rho$ which are unaffected by a unitary change of basis. Therefore, a pure state will always remain pure under unitary evolution.

\subsection{Modification of Quantum Mechanics}
It is well-known that already in ordinary quantum mechanics all measurements and the time evolution of systems can be described entirely by density matrices. Hawking proposed that a modification of quantum mechanics should be based on density matrices in order to allow a description of mixed states \cite{hawking76}. Time evolution between negative and positive infinity is then described by a superscattering operator $\slashed{S}$, so that
\[\rho_{out}=\slashed{S}\cdot\rho_{in},\]
where $\slashed{S}$ is an operator preserving hermiticity, positivity and normalisation, replacing the usual $S$ matrix. For $\rho$ representing a pure state one obtains from $\ket{\psi}_{out}=S\ket{\psi}_{in}$
\[\rho_{out}=S\rho_{in}S^{\dagger}\quad\Rightarrow\;\slashed{S}\rho=S\rho S^{\dagger}.\]
According to \cite{hawking84}, this factorisation requires the axiom of asymptotic completeness, the assumption that asymptotic states in the infinite past or future provide a basis for the full Hilbert space. \cite{hawking84} rejected the axiom for curved spacetimes and considered $\slashed{S}$ as the fundamental operator determining the dynamics of a system.
\\$\slashed{S}$ may be a non-unitary operator, but reconstruction of the initial from the final state may be possible, as $\slashed{S}$ can be invertible \cite{page}.\footnote{It was noted by Page \cite{page80} that the superscattering operator cannot be $CPT$ invariant, which means the $CPT$ theorem of quantum field theory is violated. That however relies heavily on Poincar\'e invariance.}
\\
\\Banks, Peskin and Susskind \cite{banks} now assumed that there is, at least after some coarse-grained averaging, a linear differential equation describing dynamics local in time
\[\dot{\rho}=\slashed{H}\cdot\rho.\]
It is this starting point that will have to be reconsidered later. By choosing a complete orthonormal set of hermitian matrices $Q^{\alpha}$ with $Q^0=1$, they rewrote this as\footnote{A detailed derivation can be found in the appendix.}
\[\dot{\rho}=-i[H,\rho]-\frac{1}{2}\sum_{\alpha,\beta\neq 0}h_{\alpha\beta}\left(Q^{\beta}Q^{\alpha}\rho+\rho Q^{\beta}Q^{\alpha} -2Q^{\alpha}\rho Q^{\beta}\right),\]
which is known as the {\it Lindblad equation} \cite{lindblad}. It is a Markovian master equation describing dissipative processes where quantum coherence is lost; obviously $\frac{d}{dt}(\mbox{tr }\rho)=\;\tr\dot{\rho}=0$.
\\An interesting question is which properties have to be satisfied by the couplings and operators to preserve the positivity of $\rho$; it will be discussed in section four.

\section{Attempts to Evade Information Loss}

Before a well-defined and accepted fundamental theory of physics is essentially modified, one should first try to analyse all possible alternatives to this modification. This is even more the case if this modification might lead to serious conflicts with other fundamental principles of physics.
\\John Preskill \cite{preskill} reviewed several approaches to the problem and gave arguments why all of these are no viable alternatives to information loss as reasoned by Hawking. They will be presented in this section.

\subsection{Could Information Come Out With the Radiation?}

The most obvious - and pragmatic - approach asserts that quantum effects could encode information in the outgoing radiation: The radiation could appear thermal initially, but be in fact correlated with the radiation emitted at later times. Complete knowledge of all quanta emitted during the process would then suffice to recover the initial information, and no evolution of pure into mixed states would have to occur. The analogy given by \cite{preskill,hawking05} is that of an encyclopaedia thrown into the sun; the contained information is lost in practice by all means, but a complete measurement of all radiation emitted would {\it in principle} allow reconstruction of the encyclopaedia. The goal of this approach, of course, would be to describe the complete evolution by a unitary $S$ matrix.
\\
\\There is a relatively simple argument (apparently first given by Susskind) why this viewpoint violates causality: One can draw a spacelike surface $\Sigma$ that crosses the horizon, most of the outgoing radiation, and the collapsing body well inside the horizon. Now assume the outgoing radiation is described by a pure state, so that unitarity is preserved (see fig. 2), then the initial pure quantum state evolves into a tensor product
\[\ket{i}\rightarrow\ket{i}_{in}\otimes\ket{i}_{out}.\]
But if two different states $\ket{i_1}$ and $\ket{i_2}$ evolve in this fashion, their superposition must evolve as
\[\frac{1}{\sqrt{2}}\left(\ket{i_1}+\ket{i_2}\right)\rightarrow\frac{1}{\sqrt{2}}\left(\ket{i_1}_{in}\otimes\ket{i_1}_{out}+\ket{i_2}_{in}\otimes\ket{i_2}_{out}\right),\]
which is only of the form $\ket{a}\otimes\ket{b}$ if either $\ket{i_1}_{out}$ and $\ket{i_2}_{out}$ or $\ket{i_1}_{in}$ and $\ket{i_2}_{in}$ represent the same state. Since the first case is ruled out, that means that all information gets stripped away from the infalling body as it crosses the horizon.
\begin{figure}[h]
\begin{center}
\includegraphics[scale=0.81]{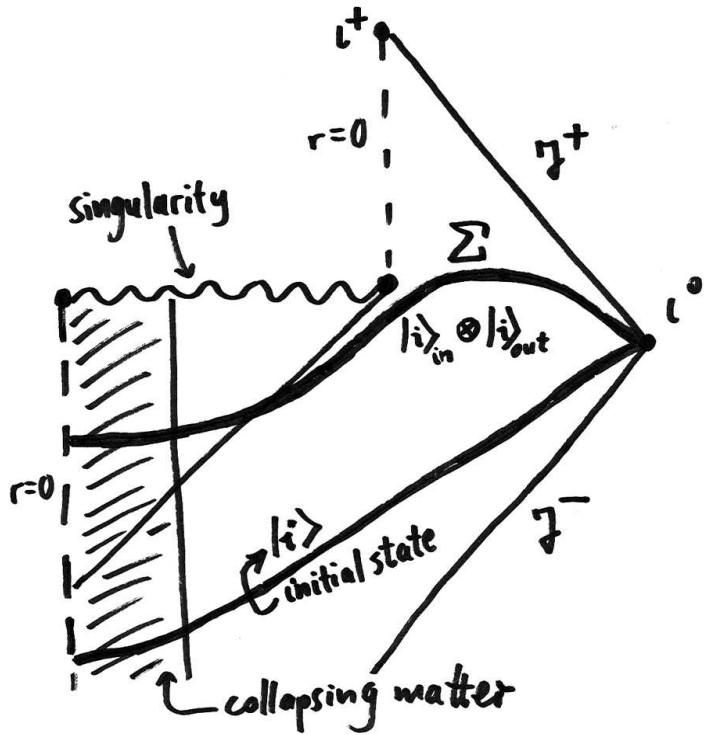}
\end{center}
\caption{If unitary evolution is assumed, the final state of the collapsing body and the outgoing radiation on $\Sigma$ must be a tensor product state.}
\end{figure}
\\
\\Indeed, the only possible way for information about a collapsing body to be encoded into radiation emitted at the event horizon that does not lead to a severe violation of causality is a mechanism that ``keeps" all information at or near the horizon.
\\Any relativist would immediately dismiss this as a violation of the equivalence principle, as an infalling observer does not observe an event horizon and nothing peculiar should happen to him.
\\
\\When presenting this argument, Page tried to argue that in quantum gravity the causal structure might be different from the classical picture, so that it may be impossible to say that $\Sigma$ is definitely spacelike (it must be nearly null classically) and a tensor product structure can not be assumed \cite{page}. But even if one accepts this argument and can explain why this encoding of information does not appear in Hawking's derivation, one still only has the choice between acausal propagation and the existence of some ``bleaching" mechanism.

\subsection{(Almost) Stable Black Hole Remnants?}

It is reasonable to expect that the semi-classical calculation breaks down when the black hole has almost radiated away, at least when it has shrunk down to Planck size. One can assume that a black hole never radiates away completely because quantum effects stop the process, so that a stable remnant of (presumably) Planck size remains. This remnant would then retain all information about the collapsing matter that formed the black hole, making it forever inaccessible. The description of the outgoing radiation by a mixed state would result from tracing over this inaccessible system.
\\
\\At first glance, this seems to be a viable possibility. However it might be hard to imagine a Planck size object containing an arbitrarily large amount of information, and it is here where problems start: The theory now contains an infinite number of different species of Planck size objects. In a process like the evaporation of a large black hole, which could be described by an ``effective field theory", there would be an amplitude for production of any of these species of Planck size objects, presumably of order of the Boltzmann factor $\exp(-\frac{M_{Pl}}{T_{BH}})$ - tiny for any large black hole, but still non-zero. If there is an infinite number of species, the total amplitude for remnant production would still be infinite, and one would expect an infinite luminosity.
It might be argued that form factors suppress the creation of black hole remnants \cite{preskill, page}, or that an ``effective field theory" description does not have to be adequate in a quantum gravity process, as Unruh and Wald pointed out in a different context \cite{unruh}. 
\\
\\There is a second, possibly more convincing, argument against the existence of such remnants: The Hawking-Bekenstein entropy $S$ of a black hole is proportional to its surface area and thus to its mass squared. The postulated black hole remnants would therefore have a small entropy, completely uncorrelated with their information content, which could be arbitrarily large. To quote \cite{preskill}, ``the beautiful edifice of black hole thermodynamics then seems like an inexplicable accident". The black hole entropy could in no way be interpreted as describing a number of internal states.
\begin{figure}[h]
\begin{center}
\includegraphics[scale=0.75]{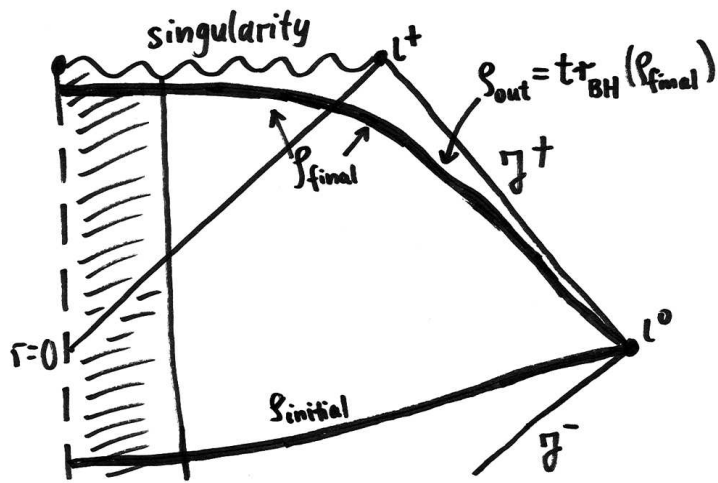}
\end{center}
\caption{If a black hole never completely evaporates, one can only consider final surfaces that cross the event horizon, and one must trace over the Hilbert space of the inaccessible interior of the black hole - just as if there was no Hawking radiation.}
\end{figure}
\\
\\Giddings proposed that a black hole might stop radiating away at a mass large compared to $M_{Pl}$, depending on its information content, so that there could be arbitrarily large stable remnants for initially arbitrarily large black holes \cite{giddings}. Although this would solve the two problems, Preskill and Page dismissed this proposal because it means that the semi-classical calculation supposedly breaks down at arbitrarily low spacetime curvature at the event horizon  \cite{preskill,page}. It does not provide a serious alternative.
\\
\\As a second, alternative possibility, it is also conceivable that all information might be encoded in the radiation emitted in the final stage of evaporation, when the black hole is of Planck size; black hole remnants would not be stable but decay. It would be natural to expect decaying rather than absolutely stable remnants, for there is no conservation law preventing their decay \cite{page}.
\\Since there is little energy left for a large amount of information to be transmitted, the remnant will take a long time to decay \cite{preskill}: If $M$ was the mass of black hole when it started evaporating, in the final stage there will be of order $S\sim M^2$ quanta needed to encode all the information. But the available energy is of order one, so every quantum will have an energy of order $M^{-2}$ and therefore a wavelength of order $M^2$. If these quanta are emitted ``one at a time", the total time will be of order $M^4$. According to \cite{page}, other authors even assumed an exponentially long time; in any case, given that the time for evaporation is of order $M^3$, there would be arbitrarily long-lived black hole remnants. One faces the same problems as if they were absolutely stable.

\subsection{Different Approaches}
The two approaches discussed so far are the alternatives to information loss most commonly found in all work on the subject. A few other ideas are also pursued:
\\
\\Firstly, anyone who believes in unitary evolution is challenged to explain how a black hole can record detailed information about any body that formed it. Hawking radiation is thermal in semi-classical theory because a black hole has no ``hair". However, Bowick et al. showed that there may be additional charges carried by a black hole, referred to as ``quantum hair" \cite{preskill,stringcrap}. These are ``axion charges" that could supposedly be ``detected by strings" in an effect analogous to the Aharonov-Bohm effect \cite{anovbohm}.
\\It is very hard to imagine that the existence of these charges might resolve the information loss paradox. As \cite{preskill} explained, one would need an infinite number of exactly conserved charges, arising from an infinite number of unbroken gauge symmetries, in order to record all information about an infalling body. This huge number of additional conservation laws should have significance for low-energy physics, where such charges have never been observed.
\\In \cite{stringcrap} the postulated existence of such charges was only used to state that a black hole can not completely evaporate since it will retain a large amount of charge for which there will be no decay channel.\footnote{The problem of remnants with too much (electrical) charge to decay was also discussed by \cite{preskill,page}; these either must be stable, causing the known problems, or conservation laws must be violated in nature.} This is just a version of the proposal of stable remnants.
\\
\\A second viewpoint was widely discussed by Page and also by Preskill \cite{preskill,page}: The Lindblad equation is commonly used to describe open quantum systems, like a system that is in contact with a thermal bath. The apparent loss of quantum coherence then stems from the fact that only part of the complete system is accessible in measurements. One can therefore view the universe as an open system from which information can leave. Indeed this is the explanation originally offered by Hawking \cite{hawking88}, who stated that ``any reasonable theory of quantum gravity will allow closed universes to branch off from our nearly flat region of spacetime." A ``superobserver" who could measure the state of the whole ``multiverse" would observe unitary evolution \cite{preskill}. This viewpoint, however, merely serves as an explanation for why evolution of pure into mixed states might be observed.
\\Page even discussed the possibility of a universe that is open in both directions, so that states can come from and go to ``baby universes". This general case would yield unpredictable mixed states. However, according to Hawking it is only possible for information to leave, not enter our universe.
\\Exciting as ideas about a ``multiverse" may seem to be, they are disheartening because they do not explain what exactly happens inside a black hole \cite{preskill}. At least they provide a picture in which information is never ``destroyed".
\begin{figure}[h]
\begin{center}
\includegraphics[scale=0.8]{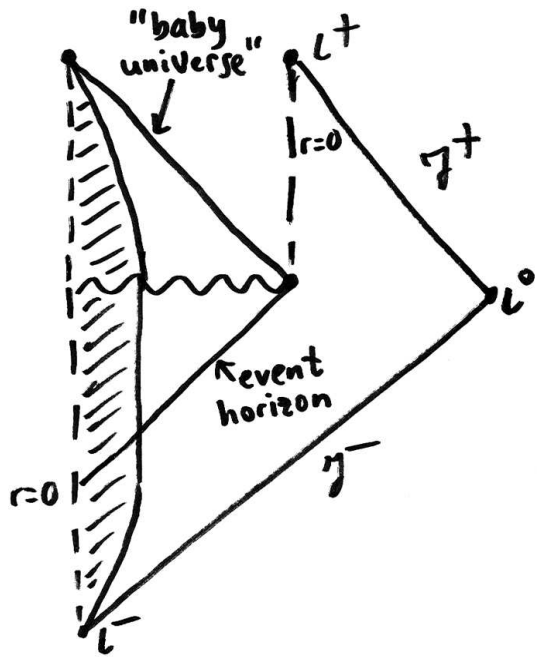}
\end{center}
\caption{A closed baby universe, causally disconnected from ours, branches off, taking away all information about the collapsing matter.}
\end{figure}
\\
\\Finally, one might also think about more drastic modifications of quantum mechanics, such as nonlinear evolution or the replacement of density matrices by something more fundamental \cite{page}. These seem to be far-fetched to most physicists; furthermore, it is unclear how these proposals could help to resolve the paradox.

\subsection{Conclusion}
Various conceivable alternatives to information loss have been scrutinised, and all of them are plagued with serious inconsistencies. If one does not believe in a force at the event horizon that strips all information away from an infalling body, assumes that the classical picture of causality for a black hole is not completely overthrown by quantum effects, and likes to attach a possible interpretation to black hole entropy, then there seems to be no way to get around information loss.
\\
\\One may choose to interpret information loss by ``baby universes" where information can go to from our universe, or one might adopt the (more radical?) viewpoint that information is simply destroyed by a black hole singularity - it seems impossible to describe a process such as the formation and evaporation of a black hole without invoking non-unitarity in quantum mechanics.

\section{Non-Unitary Laws of Physics?}

The relevant question is now whether or not it is possible to construct a sensible and consistent theory of quantum mechanics, starting from Hawking's original proposals \cite{hawking76}, that allows evolution from pure into mixed states and is not in conflict with other principles that are assumed valid. In their attempt to launch an attack on Hawking's ideas, Banks, Peskin and Susskind showed that assuming that there exists a differential equation that describes the evolution of $\rho$ which is local in time, the most general evolution equation is given by the Lindblad equation \cite{banks}. The following discussion assumes the Lindblad equation in some form as the master equation describing non-unitary physics.

\subsection{What Must Non-Unitary Laws of Physics Look Like?}
The Lindblad equation has the general form
\[\dot{\rho}=-i[H,\rho]-\frac{1}{2}\sum_{\alpha,\beta\neq 0}h_{\alpha\beta}\left(Q^{\beta}Q^{\alpha}\rho+\rho Q^{\beta}Q^{\alpha} -2Q^{\alpha}\rho Q^{\beta}\right),\]
which preserves the trace and hermiticity of $\rho$, given that $H$ and all $Q^{\alpha}$ are hermitian operators and $h_{\alpha\beta}$ is a hermitian matrix. The equation also needs to preserve positivity, i.e. the requirement that all (real) eigenvalues of $\rho$ be non-negative.
\\As shown in \cite{banks}, this is the case if $h_{\alpha\beta}$ has only non-negative eigenvalues: Diagonalise the hermitian matrix $h_{\alpha\beta}$, which can be done by a unitary transformation $u$, so that
\[h_{\alpha\beta}=\sum_{\lambda}u_{\alpha\lambda}h_{\lambda}u^*_{\beta\lambda}.\]
This means that
\[\sum_{\alpha\beta}h_{\alpha\beta}Q^{\alpha}Q^{\beta}=\sum_{\lambda}h_{\lambda}Q^{\lambda}Q^{\lambda\dagger},\]
identifying $Q^{\lambda}=\sum_{\alpha}u_{\alpha\lambda}Q^{\alpha}$. $\rho$ may also be diagonalised, so that its diagonal elements are its eigenvalues $p_i$. Now assume that one of the eigenvalues of $\rho$, say $p_1$, becomes zero; then in the evolution equation all terms but the last vanish, since $\rho_{j1}=\rho_{1j}=0$ for all $j$, so that
\[\frac{d}{dt}p_1=\dot{\rho}_{11}=\sum_{\alpha\beta}h_{\alpha\beta}(Q^{\alpha}\rho Q^{\beta})_{11}=
\sum_{\alpha\beta i}h_{\alpha\beta}(Q^{\alpha}_{1i}p_i Q^{\beta}_{i1})=
\sum_{\alpha\beta i\lambda}u_{\alpha\lambda}h_{\lambda}u^*_{\beta\lambda}(Q^{\alpha}_{1i}p_i Q^{\beta}_{i1})\]
\[=\sum_{i\lambda}h_{\lambda}\left(\sum_{\alpha} u_{\alpha\lambda}Q^{\alpha}_{1i}\right)p_i\left(\sum_{\beta} u^*_{\beta\lambda}Q^{\beta}_{i1}\right)=\sum_{i\lambda}h_{\lambda}Q^{\lambda}_{1i}p_i Q^{\lambda\dagger}_{i1}=\sum_{i\lambda}h_{\lambda}|Q_{1i}^{\lambda}|^2 p_i\ge 0\]
given that $h_{\lambda}\ge 0$ for all $\lambda$. By a similar calculation the authors of \cite{banks} also showed that entropy increases, i.e.
\[\frac{d}{dt}\;\tr(-\rho\cdot\log\rho)\ge 0,\]
if $h_{\alpha\beta}$ is a real (and therefore symmetric) matrix. However they stated that ``we do not know what conditions are necessary [...]" to have an equation that preserves positivity and increases entropy, and that one may construct examples where $h_{\alpha\beta}$ need not be real and positive. In spite of this they assumed that $h_{\alpha\beta}$ was in fact real and positive from this point on.
\\
\\A constraint on the Lindblad operators $Q^{\alpha}$ comes from the requirement of energy-momentum conservation that Hawking proposed as an additional axiom. Conservation of $H$ in the dynamics means that (assuming $H$ is time-independent) for any $k$,
\[\tr(H^k\dot{\rho})=0.\]
Now one can see that
\[\tr(H^k\dot{\rho})=\;\tr\left\{-iH^k[H,\rho]-\frac{1}{2}\sum_{\alpha,\beta\neq 0}h_{\alpha\beta}H^k\left(Q^{\beta}Q^{\alpha}\rho+\rho Q^{\beta}Q^{\alpha} -2Q^{\alpha}\rho Q^{\beta}\right)\right\}\]
\[=-\frac{1}{2}\sum_{\alpha,\beta\neq 0}h_{\alpha\beta}\;\tr\left\{Q^{\alpha}Q^{\beta}\rho H^k + Q^{\alpha}Q^{\beta}H^k\rho-2Q^{\alpha} H^k Q^{\beta}\rho\right\}.\]
As proved in \cite{liu}, if $h_{\alpha\beta}$ is real and positive, this only vanishes for general $\rho$ if $H$ commutes with all operators $Q^{\alpha}$. This is a severe constraint on the operators that could possibly appear in the Lindblad equation.
\\Srednicki concluded that the only candidates for $Q^{\alpha}$ are the Hamiltonian, the total momentum operator $\vec{P}$ or a global conserved charge \cite{srednicki}. 

\subsection{Possible Consequences of These Modifications}
Assuming certain conditions on the operators and couplings, several authors have proposed models to analyse possible observable differences from a unitary quantum theory. Several possible problems were encountered.
\\
\\First of all, Banks et al. used the above constraint on the operators $Q^{\alpha}$ already as an argument for why energy conservation is violated, as they apparently assumed that one can not satisfy the constraint \cite{banks}. They also claimed that either momentum conservation or locality would be violated by a non-unitary evolution law: A time-dependent Hamiltonian describing randomly fluctuating sources
\[H(t)=H_0+\sum_{\alpha} j_{\alpha}(t) Q^{\alpha}\]
will, in conventional quantum mechanics, lead to an evolution of the density matrix according to $\dot{\rho}(t)=-i[H(t),\rho(t)]$. Integration gives
\[\rho(\epsilon)-\rho(0) =-i\int\limits_0^{\epsilon} dt' [H_0+j_{\alpha}(t')Q^{\alpha},\rho(t')]\]
\[=-i\int\limits_0^{\epsilon} dt' [H_0+j_{\alpha}(t')Q^{\alpha},\rho(0)]-\int\limits_0^{\epsilon} dt' \int\limits_0^{t'} dt'' [H_0+j_{\alpha}(t')Q^{\alpha},[H_0+j_{\beta}(t'')Q^{\beta},\rho(0)]]+\ldots,\]
and if the sources are assumed to satisfy
\[\langle j_{\alpha}(t)j_{\beta}(t')\rangle=h_{\alpha\beta}\delta(t-t'),\]
one obtains after averaging over the sources ($\langle j_{\alpha}(t)\rangle=0$) \footnote{Note that the factor $\frac{1}{2}$ arises because of $\int\limits_0^{t'}dt''\delta(t'-t'')=\frac{1}{2}.$}
\[\frac{1}{\epsilon}\left(\rho(\epsilon)-\rho(0)\right)=-i[H_0,\rho(0)]-\frac{1}{2} h_{\alpha\beta}[Q^{\alpha},[Q^{\beta},\rho(0)]]+O(\epsilon).\]
In the limit $\epsilon\rightarrow 0$, this resembles the Lindblad equation if the matrix $h_{\alpha\beta}$ is symmetric, which is an additional requirement on which this argument crucially depends. Now Banks et al. argued that quantum mechanics with a random source breaks energy conservation. In field theory, momentum would also not be conserved if the sources were truly random because of lack of translational invariance. To preserve momentum conservation, one would need correlations between spacelike separated points with a range reciprocal to the size of momenta being added or subtracted, so that either locality or momentum conservation would have to be violated.
\\To put this more precisely, they proposed a Lindblad-type equation generalised to quantum field theory:
\[\dot{\rho}=-i\left[\int d^3 x\;H(\vec{x}),\rho\right]-\frac{1}{2}\int d^3 x\, d^3 y\;h_{\alpha\beta}(\vec{x}-\vec{y})\left(\left\{Q^{\beta}(\vec{y})Q^{\alpha}(\vec{x}),\rho\right\}-2Q^{\alpha}(\vec{x})\rho Q^{\beta}(\vec{y})\right).\]
The second term now induces nonlocal correlations including operators at different points, and the range of these correlations is given by the decay of the functions $h_{\alpha\beta}$ as $|\vec{x}-\vec{y}|$ becomes large. By expressing all quantities in terms of their Fourier transforms, one can write the second term as an integral over momentum space\footnote{The Fourier transformed  operators are no longer hermitian, so that $Q^{\dagger}$ appear because the integral must still be hermitian.}
\[-\frac{1}{2}\int \frac{d^3 p}{(2\pi)^3}\;h_{\alpha\beta}(\vec{p})\left(\left\{Q^{\dagger\beta}(-\vec{p})Q^{\alpha}(\vec{p}),\rho\right\}-2Q^{\alpha}(\vec{p})\rho Q^{\dagger\beta}(-\vec{p})\right),\]
where the operators $Q^{\alpha}(\vec{p})$ lead to violations of momentum conservation of order $|\vec{p}|$. To make this plausible, imagine that $Q(\vec{x})=\phi(\vec{x})$, where $\phi$ is a scalar field; the Fourier modes would then be just creation and annihilation operators, adding or subtracting momentum. The total magnitude of these violations is given by the decay of $h_{\alpha\beta}$ in momentum space. If $h_{\alpha\beta}$ is localized at small values of $|\vec{p}|$, there will be long-range nonlocal correlations; on the other hand, if $h_{\alpha\beta}$ falls off quickly in position space, it will be spread out in momentum space. This is reminiscent of Heisenberg's uncertainty relation which can also be viewed as a ``Fourier argument" of this type.
\\In quantum field theory, correlation functions of operators fall off at least exponentially for large spacelike separations, i.e. there are constants $C$ and $\mu$ such that at equal times
\[\left|\tr(A(\vec{x})B(\vec{y})\rho_0)\right|<Ce^{-\mu|\vec{x}-\vec{y}|},\]
where $\rho_0$ essentially represents the vacuum, cf. $\langle A(x) B(y)\rangle \equiv \braket{0|A(x)B(y)}{0}$ in quantum field theory. Banks et al. showed that while such a condition is preserved by the usual evolution law, it may be violated by a modified non-unitary evolution equation.\footnote{Details of the calculation are given in the appendix.}
\\
\\The arguments of \cite{banks} were challenged by Srednicki \cite{srednicki}: Nonlocal effects induced by non-unitarity would cause wave packets to spread out more dramatically than in conventional quantum mechanics. By explicit calculation of simple examples in quantum mechanics, such as coherent states of the harmonic oscillator, \cite{srednicki} showed that such an effect exists but it is not dramatic and decreases with time, so that no violation of locality occurs. Results in quantum field theory can be assumed to be similar. Energy and momentum conservation are guaranteed by choosing the operators $Q^{\alpha}$ appropriately as global operators commuting with the four-momentum operators $P^{\mu}\equiv(H,\vec{P})$, whereas Banks et al. had considered only local operators.
\\
\\However, Srednicki saw a different possible problem: The Lindblad equation might violate the ``weakest possible" form of Lorentz covariance, which basically means that $P^{\mu}$ transforms as a four-vector under Lorentz transformations. Assuming that all $Q^{\alpha}$ must commute with $P^{\mu}$, the fact that a product of two $Q^{\alpha}$ must transform like $H$ in order to give covariance means that $h_{\alpha\beta}$ must be purely off-diagonal (since there is no ``square root of $H$") and so have a negative eigenvalue leading to non-positivity of $\rho$.
\\This argument is not entirely watertight, since one can find examples where $h_{\alpha\beta}$ is not positive, with energy, momentum and angular momentum still conserved even though the respective operators do not commute with the $Q^{\alpha}$. Jun Liu, indeed, gave an example with these properties that also preserves positivity of $\rho$ \cite{liu}: Let $H=b^{\dagger}b$, where $b$ is a fermion operator obeying $\{b,b\}=0$ and $\{b^{\dagger},b\}=1$, then take $Q_1=b^{\dagger}+b,\;Q_2=i(b^{\dagger}-b),\;Q_3=2b^{\dagger}b$ and $h_{11}=-h_{22}=-h_{33}=g$, so that the evolution law becomes
\[\dot{\rho}=-i[H,\rho]-2g(b^{\dagger}\rho b^{\dagger}+b\rho b - b^{\dagger} b\rho -\rho b^{\dagger}b + 2b^{\dagger}b\rho b^{\dagger}b).\]
Now $\tr(H\dot{\rho})=0$ since there will always be a factor of $b^2=0$ or $(b^{\dagger})^2=0$, and energy is conserved. It becomes clear that the positivity of $h_{\alpha\beta}$ should not be viewed as a necessary condition for sensible non-unitary models.

\subsection{Observability of These Effects}
As a response to \cite{banks}, Unruh and Wald showed that violations of locality/causality or energy-momentum conservation can be kept arbitrarily small for all states that one could measure in a laboratory \cite{unruh}. Therefore it is, in principle, possible to have evolution laws that take pure states into mixed states.
\\Choosing all the $Q^{\alpha}$ in the Lindblad equation to be orthogonal projection operators and $h_{\alpha\beta}$ to be diagonal one obtains
\[\dot{\rho}=-i[H,\rho]-\sum_i\lambda_i(Q_i\rho+\rho Q_i-2Q_i\rho Q_i)=-i[H,\rho]+\sum_i\lambda_i[Q_i,[\rho,Q_i]].\]
It should be noted that $h_{\alpha\beta}$ is now real and positive, so that one would require $[H,Q_i]=0$ for energy conservation, which is not satisfied by general projection operators. This approach does not try to choose $h_{\alpha\beta}$ to be non-symmetric or indefinite to avoid problems.
\\
\\Corresponding to the Heisenberg picture of quantum mechanics, where observables evolve in time according to $\dot{A}=i[H,A]$, the evolution law is given by
\[\dot{A}=i[H,A]+\sum_i\lambda_i[Q_i,[A,Q_i]].\]
Now the argument is as follows: If $\mathcal{R}$ is an inaccessibly small region of space, e.g. of Planck size, and $R$ is a local field observable for this region, $R$ will commute with a local field observable $T$ for a disjoint region $\mathcal{T}$ at equal times:
\[[T(t),R(t)]=0.\]
Let $Q$ be a projection operator that projects on eigenstates of $R$ which have an eigenvalue greater than some given value $\alpha$; then
\[[T(t),Q(t)]=0\]
since $Q$ and $R$ are simultaneously diagonalisable. It follows that, in the evolution law where $Q$ is the only appearing projection operator,
\[\dot{T}=i[H,\rho]+\lambda[Q,[T,Q]],\]
$T$ evolves just as in conventional unitary quantum mechanics. This can be generalised to different regions $\mathcal{R}_i$ and respective projections $Q_i$, as long as $\mathcal{T}$ is disjoint from all of these. Any observable which is ``almost" global (just excludes these regions) is unaffected by the non-unitarity of the evolution law for $\rho$. Therefore the theory does not violate locality as long as one forgets about the inaccessible regions $\mathcal{R}_i$.
\\
\\Banks et al. argued that gross violations of momentum conservation would have to occur for a local theory, but \cite{unruh} showed how to confine them: One can distinguish between ``laboratory states" and ``inaccessible states", with the essential assertion that if the projection operators $Q_i$ appearing in the Lindblad equation of this model are chosen as just described, then for laboratory states $\rho_L$ one will have $Q_i\rho_L\approx Q_i\ket{0}\bra{0}$. On the scale of the regions $\mathcal{R}_i$, $\rho_L$ is essentially equal to the vacuum state. This seems reasonable if one introduces a momentum cutoff $\Lambda$ for the Hilbert space of laboratory states, for example, and then chooses $\mathcal{R}_i$ to have size much smaller than $\Lambda^{-1}$. The subspaces on which the operators $Q_i$ project depend on the parameters $\alpha_i$, the subspaces becoming smaller with increasing $\alpha_i$. Therefore $||Q_i\ket{0}||^2$ will decrease with increasing $\alpha_i$.
\\Now the values for $\lambda_i$ and $\alpha_i$ are independent parameters to adjust possible effects of non-unitarity: As loss of quantum coherence is measured by $\frac{d}{dt}\;\tr\rho^2$, the magnitude of this loss is determined by
\[\tr(\dot{\rho}\rho+\rho\dot{\rho})=\sum_i 4\lambda_i \left(\tr(Q_i\rho Q_i\rho)+\;\tr(Q_i\rho^2)\right),\]
where the first term in the Lindblad equation gives no contribution. Assuming that there are states for which the traces appearing in the sum are large, one can have rapid loss of quantum coherence by choosing the couplings $\lambda_i$ to be large. But when looking at laboratory states, the loss of quantum coherence is given by
\[\sum_i 4\lambda_i \left(\tr(Q_i\rho Q_i\rho)+\;\tr(Q_i\rho^2)\right)\approx \sum_i 4\lambda_i \left(\tr(Q_i\ket{0}\bra{0} Q_i\ket{0}\bra{0})+\;\tr(Q_i\ket{0}\bra{0})\right)\]
\[=\sum_i 4\lambda_i (||Q_i\ket{0}||^4-||Q_i\ket{0}||^2)\approx -\sum_i 4\lambda_i ||Q_i\ket{0}||^2.\]
So for any given couplings $\lambda_i$, the observable effects of non-unitary evolution, including possible violations of energy-momentum conservation, can be kept small by adjusting the operators $Q_i$. One can choose the respective values for $\alpha_i$ to be as large as necessary, so that $||Q_i\ket{0}||^2$ will be very small. Thus, all effects of non-unitary evolution will become negligible for all ``laboratory states" one might hope to create in an experiment. All problems with non-unitarity seem to be resolved as no experiment will find any effects not predicted by conventional quantum mechanics\footnote{In quantum field theory, all ``inaccessible states" will also contribute to the amplitude for a given process. I assume that these contributions will be strongly suppressed, so that no essential departure from the argument is required.}.
\\In an ``illustrative" calculation, Banks et al. had chosen the operator $Q$ to be a squared field operator. Unruh and Wald stated that this choice is inappropriate because then $||Q\ket{0}||^2$ is very large, so that the coupling $a$ (corresponding to $\lambda_i$ in \cite{unruh}) has to be chosen to be tiny. Then the apparent violation of energy conservation, proportional to $a$, will also be small. This indicates a serious loophole in the arguments of \cite{banks}.
\\
\\Ellis et al. \cite{ellis} discussed how quantum mechanics violating effects could be detected in experiments. By looking at an EPR-type situation with two particles, they showed that when a pure state evolves into a mixed state, rotational invariance no longer implies that angular momentum is conserved. Symmetries seem to no longer imply the usual conservation laws, which have to be put in by hand. This is a strong argument against non-unitary laws of physics to many \cite{page,srednicki}. In all well-accepted theories of physics, conservation laws arise naturally and do not have to be imposed.
\\However, the magnitude of any violation of conservation laws is directly related to the magnitude of the corrections to the unitary evolution law. Ellis et al. rewrote the Lindblad equation as
\[\dot{\rho}=-i[\rho,H]+\delta\slashed{H}\cdot\rho,\]
and compared the expected effects of a non-unitary term to experimental results of two separate experiments. The estimated that an upper bound for the eigenvalues of $\delta\slashed{H}$, assuming that non-unitary evolution occurs for all quantum states, should be given by
\[2\cdot 10^{-12}\;\mbox{eV},\]
obtaining the same bound independently. In light of this it must be further questioned if an apparently very small violation of conservation laws should really be regarded as unacceptable. For a macroscopic black hole the evolution of pure into mixed states might be well described even if the correction to unitary evolution is tiny, since complete evaporation takes an enormous amount of time (approx. $10^{64}$ years for a black hole of one solar mass). However, for microscopic black holes one would need rapid loss of quantum coherence, which has to be restricted to states with ``extraordinary" properties.

\subsection{A Satisfactory Resolution?}
The case for non-unitary laws of physics is still open. Assuming that some version of the Lindblad equation gives an appropriate modification of quantum mechanics, it is still not precisely known what necessary conditions the couplings $h_{\alpha\beta}$ have to satisfy to ensure positivity of $\rho$ and an increase in entropy, so that the second law of thermodynamics holds. A precise statement about which conditions are necessary would be desirable.
\\
\\A sufficient condition is that $h_{\alpha\beta}$ is a real symmetric positive matrix. One then needs operators $Q^{\alpha}$ that commute with the Hamiltonian to give conservation of energy. This severely limits the possible choices and eventually will lead to problems with Lorentz covariance. The example in \cite{liu} shows that one might get around these problems by choosing an indefinite matrix. In this case the evolution equation seems to be genuinely different. It is, however, unclear how such an evolution law might give a description of the evolution of a pure into mixed state such as to describe formation and evaporation of a black hole.
\\
\\Unruh and Wald accepted that the Lindblad equation with a real and symmetric $h_{\alpha\beta}$ will lead to violation of energy-momentum conservation and locality, but they showed how these violations may be confined to ``inaccessible" states by choosing the operators $Q^{\alpha}$ appropriately. A fundamental theory is unlikely to arise from their model, as there is no apparent reason for particular inaccessible regions $\mathcal{R}_i$ to be distinguished. 
\\Quantum mechanics can be modified in such a way that it remains local and unitary for all states that could possibly be observed in laboratory physics, but becomes non-unitary on smaller length scales. On one hand, this seems to be a promising starting point. The reason to think about non-unitary laws of physics in the first place was a phenomenon that fundamentally involves quantum gravity, and conceivably involves quantum states ``inaccessible" in laboratories. Conventional quantum mechanics would then appear as a low-energy limit. On the other hand, it appears like a cheap trick to sweep all possible problems with non-unitary quantum mechanics out of reach, and indeed, a theory that relies on ``inaccessibility in laboratory physics" is {\it a priori} untestable.
\\In black hole evaporation, non-unitary evolution would be confined to the final stage of evaporation. Before this the outgoing radiation would only appear thermal because of the trace over the Hilbert space of the inaccessible black hole interior. Then the observed von Neumann entropy would be small when the black hole has radiated away most of its mass, before sharply increasing when non-unitarity comes into play. It seems hardly possible to reconcile this with Hawking's calculation, where entropy steadily increases. 
\\
\\Two final comments should be made. Firstly, none of the authors has proposed a model to describe black hole physics. There seem to be only vague ideas of what a non-unitary theory of quantum mechanics should look like. Secondly, no argument has conclusively shown that any formulation of quantum mechanics that includes loss of information inevitably leads to inconsistencies such as violation of energy conservation or locality.

\section{New Perspectives - Back Towards Unitarity?}
In the previous sections it has been implicitly assumed that Hawking's original calculation gives an accurate picture of black hole evaporation. In the discussion of non-unitary laws of physics it was asserted that local dynamical evolution laws can describe the process. This section will turn back to these original assumptions; they could both be incorrect in a complete theory of quantum gravity. One might have to look at black hole evaporation in a slightly different way.

\subsection{Locality and Time in Quantum Gravity}
Arkani-Hamed et al. \cite{nima} pointed out that if gravity is dynamical, the notion of local observables is not a well-defined concept. One needs the metric in order to say that two points $x$ and $y$ are spacelike separated, so that one could state $[O(x),O(y)]=0$ as the requirement of locality. If that itself fluctuates, locality can not be defined precisely.
\\They interpreted this indeterminacy as an intrinsic limit on the precision of measurements for local observables: The number of internal states of any compact apparatus is bounded by the number of states of a black hole of the same size as the black hole is the object with the largest density of states. It follows that this irreducible error for a two-point correlation function is of order
\[\delta_{\langle \phi(x)\phi(y)\rangle}\sim e^{-S_{BH}}\sim e^{-\frac{|x-y|^2}{G}},\]
i.e. of order $\frac{1}{N}$, where $N$ is the number of states for a black hole that has Schwarzschild radius of order $|x-y|$. The effect is tiny and non-perturbative in $G$, but this instance of nonlocality could lead to a breakdown of Hawking's semi-classical calculation, as will be shown later.
\\
\\Secondly, the definition of time is somewhat arbitrary due to the covariance of general relativity  under diffeomorphisms (coordinate transformations), which correspond to gauge transformations. In the canonical quantisation of gravity the Wheeler-de Witt equation $H\ket{\psi}=0$ replaces the Schr\"odinger equation of quantum mechanics and contains no explicit time dependence. As shown in \cite{nima}, for non-dynamical gravity in flat space one can find a diffeomorphism invariant formulation, so that the Schr\"odinger equation is recovered. For a fluctuating metric the concept of time evolution is not well-defined.
\\
\\In \cite{hawking84}, Stephen Hawking explained why from his point of view a violation of energy-momentum conservation as described in \cite{banks} will necessarily occur if one considers dynamical laws local in space and time: \cite{hawking84} claimed that the notion of time evolution breaks down in topologically non-trivial metrics, such as on the spacetime of an evaporating black hole. For this reason the process of black hole formation and evaporation should be viewed as a scattering process where one only looks at states at positive and negative infinity, where spacetime is asymptotically flat. Energy, momentum and angular momentum are conserved globally because the initial and final states satisfy the asymptotic field equations but no conservation law should necessarily hold locally.
\\One will encounter violations of momentum conservation of order $\Delta p$ if one tries to make the dynamics local within a region $(\Delta p)^{-1}$, just as \cite{banks} claimed.
\\
\\From this perspective there is no reason to assume that local evolution laws describe evolution from pure states into mixed states. It seems that not only can quantum gravity not be described by conventional quantum field theory, but also that the dynamics of quantum gravity seem to be very different from those of known theories of physics.
\begin{figure}[h]
\begin{center}
\includegraphics[scale=0.67]{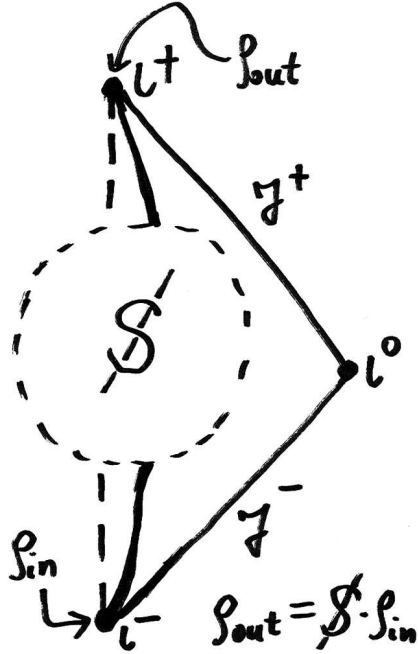}
\end{center}
\caption{In quantum gravity, measurements are made where spacetime is asymptotically flat, without a local description of the evolution of an initial into a final state.}
\end{figure}

\subsection{Black Hole Complementarity}
An intriguing new proposal to get around the usual arguments against unitary evolution in black-hole evaporation was provided by Susskind, Thorlacius, and Uglum \cite{complement}. This is based on the earlier observation by Thorne, Damour and others that a classical black hole can be described in terms of a so-called ``stretched horizon", which has the physical (thermal, electrical etc.) properties of a membrane. This may be confusing, as it was argued earlier that nothing peculiar should happen at the event horizon of a black hole, or anywhere but close to the curvature singularity. Indeed a freely falling observer will not perceive a membrane, but a stationary observer remaining outside of the black hole will. This leads to a notion of {\it black hole complementarity}, expressed in the following three postulates:
\begin{itemize}
\item The process of black hole formation and evaporation can be described within standard quantum theory, and is in particular unitary.
\item The semi-classical approximation is a good approximation to physics outside of the stretched horizon.
\item The entropy $S$ does describe microscopic degrees of freedom of the black hole, as observed by a distant observer.
\end{itemize}
This proposal could be regarded as the ``maximally conservative" proposed solution to the problem, concerned only with observations made from outside the black hole, and may not sound very satisfactory. It was demonstrated in \cite{complement}, however, how to make sense of it, and in particular of the role of the stretched horizon, in a simplified model of two-dimensional ``dilaton" (scalar) gravity. The physical validity of semi-classical arguments of the analysis was put in question by Hawking \cite{dilaton}, who argued that because the temperature and rate of emission of radiation of an evaporating black hole in this model remain finite the evaporation process must lead to a naked singularity if the approximation is trusted. In any case it is probably fair to say that a simplified two-dimensional model can not give more than indications that one is on the right track.
\\
\\Let us assume black hole complementarity as a guiding principle. What about the apparently inevitable conclusions of section three, then? According to black hole complementarity, it is not meaningful to speak of a tensor product Hilbert space of states outside and inside of the event horizon. Only a nonexistent ``superobserver" having access to regions both inside and outside of the horizon could make measurements on a state in this space, or make any predictions about its time evolution. One does not try to describe the universe as a whole in one consistent quantum theory, but only demands that time-evolution is consistent from the perspective of all possible observers. As noted in \cite{complement}, this may well mean that the present framework of quantum field theory is inadequate to describe black-hole evaporation.
\\
\\Imposing such a principle, of course, does not give any indication of a dynamical process that might encode information in the outgoing radiation. It still seems that all information can only come out ``at the end", and this leads to the problems outlined in section three, as this process would be assumed to take a very long time. Building on black hole complementarity, Hayden and Preskill \cite{haydpres} described a simple quantum information-theoretic model of how information could become scrambled in a black hole. Their proposal was not meant to provide new conceptual insights, but showed how simple estimates might be misleading because one has to ``classical" a picture in mind. Firstly, they assumed that, for a black hole composed of $n$ qubits, an instantaneous, randomly chosen unitary transformation ``randomizes" this information before the qubits get released in Hawking radiation. The black hole works as what is known as a {\it quantum erasure channel}, which has the property that after $(n+k+c)/2$ qubits have been released, an arbitrary $k$ qubits of information can be retrieved with fidelity at least $1-2^{-c}$. Thus, for an arbitrarily large black hole one can get arbitrarily close to ultimately retrieving all of the information thrown into the black hole (keeping $c$ constant). As soon as the black hole has evaporated more than half of its original information content, information can be retrieved from the radiation.
\\
\\This, of course, is a very simplified model, and even if one assumes black hole complementarity, leads to a contradiction: Assume a black hole has already evaporated away more than half of its original information, and most of the outgoing radiation has been received by an observer outside of the black hole. A second observer could then fall into the black hole, sending off a message about his information before becoming almost instantly ``randomized". The first observer, having retrieved this information through the following Hawking radiation, could then follow into the black hole, receive the message and thus achieve some kind of quantum ``cloning", which is not possible (see Fig. 6). This would violate the postulate that standard quantum mechanics can describe the process.
\begin{figure}[h]
\begin{center}
\includegraphics[scale=0.67]{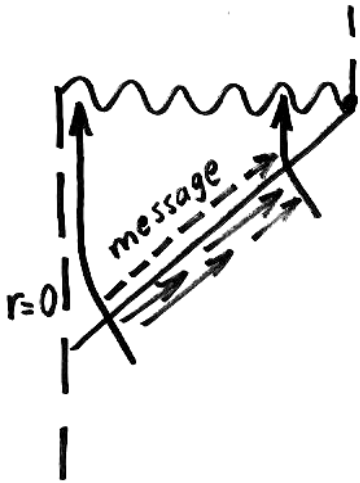}
\end{center}
\caption{If infalling information gets randomized and encoded in outgoing radiation almost immediately, cloning of a quantum state is possible.}
\end{figure}
\\To test if this problem is still present in a more realistic model, an estimate for the ``thermalization time" of a black hole was given in \cite{haydpres}. One would normally assume that a perturbation of a black hole dies off within a time scale $M$, since this is true classically, but only locally. Using the dissipative properties that the proposed stretched horizon has, one obtains an estimate of $M \log M$ (measured in the Schwarzschild time coordinate) for the black hole to settle down to a quasi-stationary state, which was shown can be achieved by a model of efficient quantum circuits.
\\One can then show \cite{haydpres} that the proper time that the infalling observer has to send a message to the other observer is of order $M e^{-\Delta t/M}$, where $\Delta t$ is the difference in Schwarzschild time between the two observers' crossing the horizon. Since this must be at least equal to the thermalization time for the external observer to retrieve information while he is still outside of the horizon, the proper time left to send the message is at most of the order of the Planck time, and in particular cannot be increased by choosing $M$ to be large. Since this means super-Planckian frequencies are required, one would normally regard this as impossible, and cloning is (``just barely") prevented.
\\
\\From a quantum information theoretic perspective, such a randomization of the infalling information also means that no acausal signalling occurs in this process, and no mysterious ``bleaching" mechanism seems to be required. While the proposed model of Hayden and Preskill may look overly simplified to give insights into quantum gravity, it shows that information-theoretic reasoning can sometimes lead to very different conclusions than naively expected. It appears that the objections of section three could, at least in principle, be avoided. It was advocated in \cite{haydpres} there there could be ``an interesting middle ground" between unitarity and information loss - the interpretation of this is somewhat unclear.

\subsection{Holography and String Theoretic Arguments}
Most string theorists believed that the AdS/CFT correspondence\footnote{supposedly ``the greatest advance in theoretical physics over the last ten years" \cite{tong}}, proposed by Maldacena \cite{maldacena}, solved the information loss paradox, and that no information was lost \cite{hawking05,susskind}. Put simply, according to the correspondence, all gravitational processes can be mapped to a conformal field theory on the boundary, which is unitary. It seems that to trust these ideas one has to accept string theory as a consistent theory of quantum gravity.
\\
\\However, the AdS/CFT correspondence is a realisation of a more general principle. The result for the Hawking-Bekenstein entropy, namely that the entropy of a black hole is proportional to its surface area and not its volume, had led to the conjecture of the ``holographic principle" which states that a region with boundary of area $A$ is fully described by no more than $\frac{A}{4}$ degrees of freedom \cite{bousso}, since black holes have maximal entropy. This conclusion does not seem to depend on string theory or supersymmetric field theory. It means that a local field theory in which the number of degrees of freedom is proportional to the volume may not give a sufficent description when gravity is involved. In the context of black hole evaporation, the assumption that time evolution is unitary means that the entropy of the outgoing Hawking radiation is entanglement entropy which can not exceed the entropy of the system that forms the second subsystem, i.e. the black hole \cite{nima}. At the point where the entropy of the remaining black hole becomes equal to the total entropy of the emitted radiation, the semi-classical approximation has to break down and the observed entropy of the radiation must decrease and finally become zero.
\\
\\At this point the black hole is still large compared to the Planck scale. \cite{nima} gives an idea of how one could explain this apparently early breakdown of the semi-classical calculation: Consider $N$ spins $\sigma_i=\pm\frac{1}{2}$ and the pure state
\[\ket{\psi}=\sum_{\sigma_i}\frac{1}{2^{\frac{N}{2}}}\ket{\sigma_1\ldots\sigma_N}e^{i\theta(\sigma_1,\ldots,\sigma_N)},\]
where $\theta(\sigma_1,\ldots,\sigma_N)$ are random phases. If one only measures $k$ spins, the state appears as a mixed state, with entropy resulting from tracing over the unobserved subsystem
\[S=k\log 2 + O(2^{-N+2k}).\]
The correction is exponentially small in $N$, but also exponentially increasing with the number of measurements, so that the effect is of order one when $k\approx\frac{N}{2}$. For $k=N$ one will have entropy zero, since $\ket{\psi}$ is a pure state.
\\Likewise, exponentially small corrections to the leading order semi-classical theory can add up and significantly contribute when the number of emitted quanta of Hawking radiation is of order $S_{BH}$, which is the case after a time scale of the order of the (semi-classical) evaporation time $t\sim S_{BH}^{3/2}$. Just as in the spin case, the observed entropy will, after this time, drop to zero, so that unitarity is ultimately restored.
\\Though an explanation for the breakdown of semi-classical theory at low curvature might emerge from these ideas, the apparent inconsistencies for any picture of unitary evolution explained in section three still exist. Again one has to invoke black hole complementarity. While an outside observer indeed observes ``bleaching" at the horizon, the infalling observer will not observe anything special. Certainly, a departure from the picture of causality offered by classical general relativity is required.
\\
\\It is usually assumed that Hawking radiation is a quantum-gravitational prediction, and must therefore be reproduced, and perhaps explained, by any theory of quantum gravity. It was therefore considered to be a great success of string theory when Strominger and Vafa gave a calculation \cite{strominger} which could be interpreted as explaining black-hole entropy in terms of microscopic degrees of freedom, and string theorists claimed this puzzle had been solved. In the words of Roger Penrose \cite{penrose}, however, ``as appears to be usual, with such string-theoretic proclamations, this conclusion is very considerably overblown". This is because all calculations are performed in flat space, and the horizon of a black hole seems to not have played a role in these calculations, which could only be performed for near-extremal black holes with positive specific heat (as opposed to the usual {\it negative} specific heat). It must be noted, however, that the exact agreement with the Hawking-Bekenstein entropy is quite remarkable, and at least gives evidence to the idea of the holographic principle.
\\Similar calculations in loop quantum gravity \cite{lqg}, while giving a clearer ``picture" and appearing to be better motivated physically, gave an entropy proportional to the area of the black hole, while the prefactor depended on the Barbero-Immirzi parameter (a free parameter of the theory). Again, the prediction of a universal formula for similar types of black holes seems quite remarkable, but does not seem to provide an intuitive understanding of the microscopic degrees of freedom of the black hole, or of any physical mechanism ``destroying" or ``encoding" information.

\subsection{No Need For Non-Unitarity, After All?}
In 2004, Stephen Hawking claimed that he had found an explanation why no evolution from pure into mixed states should occur in quantum gravity \cite{hawking05}. The basic statement is that since all measurements can be made only at spatial infinity, one does not know if a black hole has formed and evaporated in the process. One has to take the sum over all possible histories in the path integral, which include those in which there was no black hole. 
\\This sum restores unitarity for the following reasons: When calculating the partition function of gravity\footnote{For a brief account of the path integral approach to quantum gravity, the reader is referred to the appendix.}, one takes the Euclidean path integral (``the only sane way to do quantum gravity nonperturbatively" \cite{hawking05}) over metrics of all topologies that fit inside the boundary $S^2\times S^1$. The point is that the path integral over metrics with trivial topology is unitary, while the path integral over metrics with non-trivial topology (including black holes) gives correlation functions that decay to zero (``very plausible" \cite{hawking05}), so that they will not contribute at late times, even though classically there might have been a very high probability for a black hole to form. In the end, one obtains a unitary mapping from initial to final states. In a semi-classical approximation one ``throws away unitarity."
\\
\\If the arguments of \cite{hawking05} are accepted (though there seems to be no proof and they seem to apply only to spacetimes which are asymptotically AdS) the lesson still remains highly elusive. What happens when a black hole forms and evaporates? According to \cite{hawking05}, Hawking and Hartle showed in \cite{hartle} that radiation could be thought of as tunnelling through the event horizon, and this might explain how it could carry information. But the final viewpoint seems to require that anything that happens between negative and positive infinity can not be specified. One has to take all possibilities into account, just as in microscopic processes described by conventional quantum field theory.
\\
\\\cite{hawking05} contained few explicit calculations, and it seems hard to compare the results to the semi-classical approach and see what went wrong there. The original question if black hole evaporation implies that physics is non-unitary is more open than ever.

\section{Conclusion/Outlook}
The title of this essay asked two questions. The first one already seemed to be answered: Hawking radiation suggests that physics is non-unitary, and it seemed to have been shown conclusively that there is no alternative to this conclusion, unless one wants to make a great sacrifice such as disregarding locality to preserve unitarity.
\\But in the end, it looks as if a different perspective on quantum gravity might be required. Were all examined alternatives to information loss just ``classical" pictures inadequate to describe a process presumably including physics at the Planck scale? Will it prove to be impossible to describe quantum gravity by a ``local" theory? That would eventually mean that a departure from the present understanding of cause and effect will be required in this theory, as the holographic principle suggests.
\\
\\These questions will presumably remain open for the foreseeable future. Assuming for the moment that there is information loss in black holes and that a description of it by a local theory is meaningful, the second question is: Can the present laws of physics be modified to accommodate information loss? Though nobody seems to have an idea of how exactly this should be done, the answer seems to be yes. No cherished principle is necessarily violated if evolution of pure into mixed states is possible.
\\Inapt choices for the quantities remaining in the Lindblad equation will raise potential difficulties. But it seems fair to say that a consistent theory can be based on some version of the Lindblad equation, so that evolution of pure states into mixed states occurs. At least a little insight has been gained into what non-unitary laws of physics must be like.
\\It may be necessary, as in Unruh's and Wald's model, to confine non-unitary evolution to a domain where it can escape present experiments, but become important at small length scales. However, assuming that non-unitarity only occurs at the Planck scale will presumably not allow the construction of a consistent theory that accurately describes processes involving black holes.
\\Any hope for experimental data seems far out at present. Maybe the physics world will be surprised by an observation of non-unitary evolution in some future experiment. Accordingly one can hope to give more definite statements based on experimental results, or rely on pure thought in the meantime. But, to quote John Preskill, ``anyway, we don't have much choice."

\def\thesection{\Roman{section}}
\setcounter{section}{9}

\section{Appendix}

\subsection{Derivation of the Lindblad Equation}
Starting from
\[\dot{\rho}=\slashed{H}\cdot\rho,\]
where $\slashed{H}$ is linear and the right-hand side must be hermitian, this can be written as
\[\dot{\rho}=-\sum_{\alpha\beta}h_{\alpha\beta}Q^{\alpha}\rho Q^{\beta},\]
where $Q^{\alpha}$ is an arbitrary complete orthonormal set of hermitian matrices with $Q^0=1$ and $h_{\alpha\beta}$ is a hermitian matrix. The requirement that $\tr\rho$ must be constant is the constraint
\[0=\;\tr\left(h_{00}\rho+\sum_{\alpha\neq 0}(h_{0\alpha}+h_{\alpha 0})Q^{\alpha}\rho+\sum_{\alpha,\beta\neq 0}h_{\alpha\beta}Q^{\beta}Q^{\alpha}\rho\right),\]
using the cyclic property of the trace. This can be satisfied by $h_{00}=0$ and 
\[\sum_{\alpha,\beta\neq 0}h_{\alpha\beta}Q^{\beta}Q^{\alpha}=-\sum_{\alpha\neq 0}(h_{0\alpha}+h_{\alpha 0})Q^{\alpha}.\]
Since the $Q^{\alpha}$ form a complete set, $Q^{\beta}Q^{\alpha}=\sum_{\gamma\neq 0} g_{\beta\alpha\gamma}Q^{\gamma}$ for some $g_{\beta\alpha\gamma}$, so that $(h_{0\alpha}+h_{\alpha 0})=-\sum_{\beta,\gamma\neq 0} h_{\beta\gamma}g_{\gamma\beta\alpha}$ satisfies the condition. After the parametrisation
\[\sum_{\alpha\neq 0}(h_{0\alpha}-h_{\alpha 0})Q^{\alpha}=2iH,\]
which defines a hermitian operator $H$ since $h_{\alpha\beta}$ is hermitian and so the left-hand side is anti-hermitian, one obtains
\[\sum_{\alpha\neq 0}h_{0\alpha}\rho Q^{\alpha}=\sum_{\alpha\neq 0}\frac{1}{2}\left(-\sum_{\beta,\gamma\neq 0}h_{\beta\gamma}g_{\gamma\beta\alpha}+(h_{0\alpha}-h_{\alpha 0})\right)\rho Q^{\alpha}\]
\[=-\frac{1}{2}\rho\sum_{\alpha,\beta,\gamma\neq 0}h_{\beta\gamma}g_{\gamma\beta\alpha}Q^{\alpha}+i\rho H=-\frac{1}{2}\rho\sum_{\beta,\gamma\neq 0}h_{\beta\gamma}Q^{\gamma}Q^{\beta}+i\rho H\]
and similarly
\[\sum_{\alpha\neq 0}h_{\alpha 0}Q^{\alpha} \rho=-\frac{1}{2}\sum_{\beta,\gamma\neq 0}h_{\beta\gamma}Q^{\gamma}Q^{\beta}\rho-iH\rho, \]
which gives the most general form for a differential equation preserving $\tr\rho$
\[\dot{\rho}=-i[H,\rho]-\frac{1}{2}\sum_{\alpha,\beta\neq 0}h_{\alpha\beta}\left(Q^{\beta}Q^{\alpha}\rho+\rho Q^{\beta}Q^{\alpha} -2Q^{\alpha}\rho Q^{\beta}\right),\]
the Lindblad equation.

\subsection{Possible Violation of Locality For Generalised Lindblad-Type Equation}
Following \cite{banks}, assume that for some given density matrix $\rho_0$ the following holds for sufficiently large separations $|\vec{x}-\vec{y}|$ for all local operators
\[|\tr(A(\vec{x})B(\vec{y})\rho_0)|<Ce^{-\mu|\vec{x}-\vec{y}|}.\]
In standard quantum mechanics the evolution law gives
\begin{eqnarray*}
\frac{d}{dt}\tr(A(\vec{x})B(\vec{y})\rho)\Big|_{t=0} & = & i\,\tr\left(A(\vec{x})B(\vec{y})[H,\rho_0] \right)
\\& = & -i\,\tr\left(([A(\vec{x}),H]B(\vec{y})+A(\vec{x})[B(\vec{y}),H])\rho_0\right),
\end{eqnarray*}
and since both commutators are local operators, both terms will be exponentially small, so that
\[\left|\frac{d}{dt}\tr(A(\vec{x})B(\vec{y})\rho_0)\right|<2Ce^{-\mu|\vec{x}-\vec{y}|}.\]
There can be a spread of wave packets, so that the left-hand side is non-zero and correlations develop over time, but there will be an exponential drop-off of correlation functions at all times. Now the second term in the generalised equation asserted in \cite{banks} will give an additional contribution, given by (where from now on $A\equiv A(\vec{x})$ and $B\equiv B(\vec{y})$)
\begin{eqnarray*}
K(A,B) & := & -\frac{1}{2}\,\tr\int d^3 z\; d^3 w\; AB\;h_{\alpha\beta}(\vec{z}-\vec{w})\left(\left\{Q^{\beta}(\vec{w})Q^{\alpha}(\vec{z}),\rho_0\right\}-2Q^{\alpha}(\vec{z})\rho_0 Q^{\beta}(\vec{w})\right)
\\&=&-\frac{1}{2}\,\tr\int d^3 z\; d^3 w \;h_{\alpha\beta}(\vec{z}-\vec{w})\left(\left[AB,Q^{\beta}(\vec{w})\right]Q^{\alpha}(\vec{z})+Q^{\beta}(\vec{w})\left[Q^{\alpha}(\vec{z}),AB\right] \right)\rho_0
\end{eqnarray*}
Now expanding the commutators, this is equal to
\[-\frac{1}{2}\,\tr\left\{\int d^3 z\; d^3 w \;h_{\alpha\beta}(\vec{z}-\vec{w})\left(\left[A(\vec{x}),Q^{\beta}(\vec{w})\right]B(\vec{y})Q^{\alpha}(\vec{z})+A(\vec{x})\times\right.\right.\]
\[\left.\left.\times\left[B(\vec{y}),Q^{\beta}(\vec{w})\right]Q^{\alpha}(\vec{z})-Q^{\beta}(\vec{w})A(\vec{x})\left[B(\vec{y}),Q^{\alpha}(\vec{z})\right]-Q^{\beta}(\vec{w})\left[A(\vec{x}),Q^{\alpha}(\vec{z})\right]B(\vec{y}) \right)\rho_0\right\}\]
A commutator like $\left[A(\vec{x}),Q^{\beta}(\vec{w})\right]$ will only be non-zero for $\vec{w}=\vec{x}$ and so 
\[\int d^3 z\;h_{\alpha\beta}(\vec{z}-\vec{w})\left[A(\vec{x}),Q^{\beta}(\vec{w})\right]=\int d^3 z\;h_{\alpha\beta}(\vec{z}-\vec{x})\left[A(\vec{x}),Q^{\beta}(\vec{w})\right].\]
Use this to rewrite
\begin{eqnarray*}
K(A,B) & =&-\frac{1}{2}\,\tr\left\{\int d^3 z \;h_{\alpha\beta}(\vec{z}-\vec{x})\left(\left[A(\vec{x}),\int d^3 w\; Q^{\beta}(\vec{w})\right]B(\vec{y})Q^{\alpha}(\vec{z})\right.\right.
\\ &  &+\left.A(\vec{x})\left[B(\vec{y}),\int d^3 w\; Q^{\beta}(\vec{w})\right]Q^{\alpha}(\vec{z})\right)\rho_0-\int d^3 w\; h_{\alpha\beta}(\vec{x}-\vec{w}) \left(Q^{\beta}(\vec{w}) \times\right.
\\ & & \left.\left. \times A(\vec{x})\left[B(\vec{y}),\int d^3 z\; Q^{\alpha}(\vec{z})\right]-Q^{\beta}(\vec{w})\left[A(\vec{x}),\int d^3 z\; Q^{\alpha}(\vec{z})\right]B(\vec{y}) \right)\rho_0\right\},
\end{eqnarray*}
which, using the commutativity of spacelike separated operators, so that for example $[[A(\vec{x}),\int\,Q^{\beta}],B(\vec{y})]=0$, can be written more succinctly as
\begin{eqnarray*}
K(A,B) &=&-\frac{1}{2}\,\tr\left\{\left(\int d^3 z \;h_{\alpha\beta}(\vec{z}-\vec{x})\left[A(\vec{x}),\int d^3 w\; Q^{\beta}(\vec{w})\right]B(\vec{y})Q^{\alpha}(\vec{z})\right.\right.
\\&&\left.\left.-\int d^3 w\; h_{\alpha\beta}(\vec{x}-\vec{w}) Q^{\beta}(\vec{w})B(\vec{y})\left[A(\vec{x}),\int d^3 z\; Q^{\alpha}(\vec{z})\right]\right)\rho_0\right\}+(A\leftrightarrow B).
\end{eqnarray*}
Now a density matrix $\rho$ has the property that for localised operators $C(\vec{x})$ and $D(\vec{y})$,
\[\tr(C(\vec{x})D(\vec{y})\rho)\approx\tr(C(\vec{x})\rho)\tr(D(\vec{y})\rho)\]
for large separations $|\vec{x}-\vec{y}|$ \cite{srednicki}. This can be used to split the traces into two factors:
\begin{eqnarray*}
K(A,B)&=&-\frac{1}{2}\int d^3 z \;h_{\alpha\beta}(\vec{z}-\vec{x})\,\tr\left\{\left(\left[A,\int d^3 w\; Q^{\beta}(\vec{w})\right]B Q^{\alpha}(\vec{z})\right)\rho_0\right\}
\\&&+\frac{1}{2}\int d^3 w\; h_{\alpha\beta}(\vec{x}-\vec{w})\,\tr\left\{ \left(Q^{\beta}(\vec{w})B\left[A,\int d^3 z\; Q^{\alpha}(\vec{z})\right]\right)\rho_0\right\}
\\&&+(A(\vec{x})\leftrightarrow B(\vec{y})).
\\&\approx& -\frac{1}{2}\,\tr\left\{\left[A,\int d^3 w\; Q^{\beta}(\vec{w})\right]\rho_0\right\}\int d^3 z \;h_{\alpha\beta}(\vec{z}-\vec{x})\,\tr\left\{B Q^{\alpha}(\vec{z})\rho_0\right\}
\\&&+\frac{1}{2}\,\tr\left\{\left[A,\int d^3 z\; Q^{\alpha}(\vec{z})\right]\rho_0\right\}\int d^3 w\; h_{\alpha\beta}(\vec{x}-\vec{w})\,\tr\left\{ Q^{\beta}(\vec{w})B\rho_0\right\}
\\&&+(A(\vec{x})\leftrightarrow B(\vec{y})).
\end{eqnarray*}
By relabelling $((\vec{z},\alpha)\leftrightarrow(\vec{w},\beta))$ in the second term and using the hermiticity of $h_{\alpha\beta}$ this can be cast in the form
\begin{eqnarray*}
K(A,B)&\approx& -\frac{1}{2}\,\tr\left\{\left[A,\int d^3 w\; Q^{\beta}(\vec{w})\right]\rho_0\right\}\int d^3 z \;h_{\alpha\beta}(\vec{z}-\vec{x})\,\tr\left\{B Q^{\alpha}(\vec{z})\rho_0\right\}
\\&&+\frac{1}{2}\,\tr\left\{\left[A,\int d^3 w\; Q^{\beta}(\vec{w})\right]\rho_0\right\}\int d^3 z\; h^*_{\alpha\beta}(\vec{z}-\vec{x})\,\tr\left\{ Q^{\alpha}(\vec{w})B\rho_0\right\}
\\&&+(A(\vec{x})\leftrightarrow B(\vec{y})).
\end{eqnarray*}
Splitting $h_{\alpha\beta}=\Re h_{\alpha\beta}+i\Im h_{\alpha\beta}$ one finally obtains
\begin{eqnarray*}
K(A,B)&\approx& -\frac{1}{2}\,\tr\left\{\left[A(\vec{x}),\int Q^{\beta}\right]\rho_0\right\}\int d^3 z \;\Re\,h_{\alpha\beta}(\vec{z}-\vec{x})\,\tr\left\{\left[B(\vec{y}), Q^{\alpha}(\vec{z})\right]\rho_0\right\}
\\&&-\frac{i}{2}\,\tr\left\{\left[A(\vec{x}),\int Q^{\beta}\right]\rho_0\right\}\int d^3 z \;\Im\, h_{\alpha\beta}(\vec{z}-\vec{x})\,\tr\left\{\left\{B(\vec{y}), Q^{\alpha}(\vec{z})\right\}\rho_0\right\}
\\&&+(A(\vec{x})\leftrightarrow B(\vec{y}))
\\&=& -\frac{1}{2}\,\tr\left\{\left[A(\vec{x}),\int Q^{\beta}\right]\rho_0\right\}\Re\,h_{\alpha\beta}(\vec{y}-\vec{x})\,\tr\left\{\left[B(\vec{y}), \int d^3 z \; Q^{\alpha}(\vec{z})\right]\rho_0\right\}
\\&&-\frac{i}{2}\,\tr\left\{\left[A(\vec{x}),\int Q^{\beta}\right]\rho_0\right\}\int d^3 z \;\Im\, h_{\alpha\beta}(\vec{z}-\vec{x})\,\tr\left\{\left\{B(\vec{y}), Q^{\alpha}(\vec{z})\right\}\rho_0\right\}
\\&&+(A(\vec{x})\leftrightarrow B(\vec{y})),
\end{eqnarray*}
and this is essentially the result given in \cite{banks}. In the last step the fact that $\left[B, Q^{\alpha}(\vec{z})\right]=0$ unless $\vec{y}=\vec{z}$ was used again.
\\Now since one may choose any local operators $A(\vec{x})$ and $B(\vec{y})$ in this expression, the appearing traces should be expected to be possibly large. Then the nonlocal correlations that develop over time depend on the drop-off of the function $h_{\alpha\beta}(\vec{x}-\vec{y})$, as was already claimed on rather heuristic grounds before.

\subsection{Path Integral Approach to Quantum Gravity}
This introduction follows \cite{hawking78}. In conventional quantum field theory, the path integral approach invented by Feynman gives a nonperturbative description of any field theory and has proved to be a powerful mathematical tool. One may therefore attempt to quantise gravity by working out the respective path integral. The amplitude to go from an initial state at time $t$ with metric $g$ and matter fields $\phi$ to a state at time $t'$ with metric $g'$ and matter fields $\phi'$ would be given by
\[\langle g',\phi',t'|g,\phi,t\rangle=\int d[g]\;d[\phi]\;e^{iI[g,\phi]},\]
where the gravitational part of $I$ is
\[I=\frac{1}{16\pi G}\int_{\mathcal{M}} d^4 x\;\sqrt{g}\;R+\frac{1}{8\pi G}\int_{\mathcal{\partial M}} d^3 x\;\sqrt{h}\;K+C[h],\]
where $\mathcal{M}$ is the spacetime manifold and one will choose $C[h]$ so that $I=0$ for Minkowski space. A partition function for the canonical ensemble at temperature $T=\frac{1}{\beta}$ will be given by
\[Z=\tr(\exp(-\beta H[g_L]))=\int d[g]\;d[\phi]\;e^{-I[g_R,\phi]}\]
where in the last step a Wick rotation was performed, so that $-\beta H[g_L]=\beta L[g_R]=-I[g_R]$ because in a thermal background imaginary time is periodic with period $\beta$, where $g_L$ is the Lorentzian and $g_R$ is the Riemannian metric. The path integral is then taken over all metrics whose boundary is a two-sphere at infinity times a circle of circumference $\beta$, representing periodicity in time, or in other words, the topology at infinity is $S^2\times S^1$.
\\One will typically try to expand this path integral about a classical solution by setting
\[I[g,\phi]=I[g_0,\phi_0]+I_2[\tilde{g},\tilde{\phi}]+\ldots,\]
where $I_2$ is quadratic in the perturbations $\tilde{g}$ and $\tilde{\phi}$, so that
\[Z\approx e^{-I[g_0,\phi_0]}\int d[\tilde{g}]\;d[\tilde{\phi}]\;e^{-I_2[\tilde{g},\tilde{\phi}]}.\]
In the case where the background metric is a Schwarzschild black hole, it will contribute to the partition function at $\beta=8\pi M$. Only the surface term contributes to the action $I=\frac{\beta^2}{16\pi}$. To first order the partition function will then be
\[Z\approx e^{-\frac{\beta^2}{16\pi}},\]
from which one can obtain the expectation value for the energy
\[\langle E\rangle=-\frac{d}{d\beta}(\log Z)=\frac{\beta}{8\pi}=M,\]
which gives the correct result for the Hawking-Bekenstein entropy
\[S=\beta\langle E\rangle +\log Z=8\pi M^2 - 4\pi M^2 = 4\pi M^2=\frac{A}{4}.\]

\section*{Acknowledgements}
I would like to thank Jonathan Oppenheim for supervising this essay and making me think about these interesting things, and Philip Tanedo for pointing out mistakes and inaccuracies in my English. I am also grateful to the referee for helpful comments on section 5 and the introduction.

\end{document}